\documentclass[oneside,reqno]{amsart}
\usepackage{amsmath, amsthm, mathrsfs, amssymb, amsfonts, mathtools}
\usepackage{graphicx}
\graphicspath{ {Figures/} }
\usepackage{hyperref}
\hypersetup{colorlinks=true, linkcolor=blue, citecolor=magenta, filecolor=magenta, urlcolor=cyan}
\usepackage{booktabs}
\usepackage[backend=biber, style=phys, doi=false, isbn=false, hyperref=true, backref=false, giveninits=true, abbreviate=true, sorting=nyt]{biblatex}
\bibliography{lib}

\numberwithin{equation}{section}
\mathtoolsset{showonlyrefs}

\usepackage[colorinlistoftodos, textwidth=1.75cm]{todonotes}
\reversemarginpar

\usepackage{lmodern}
\makeatletter
\ifcase \@ptsize \relax
  \newcommand{\miniscule}{\@setfontsize\miniscule{4}{5}}
\or
  \newcommand{\miniscule}{\@setfontsize\miniscule{5}{6}}
\or
  \newcommand{\miniscule}{\@setfontsize\miniscule{5}{6}}
\fi
\makeatother

\newcounter{todocounter}

\def\diff{\mathrm{d}}

\def\1{\mathbb{1}} 

\newcommand{\editremove}[1]{}
\newcommand{\edittrash}[1]{}

\newcommand{\editcross}[1]{\xout{#1}}

\renewcommand{\editcross}[1]{}

\usepackage[foot]{amsaddr}

\begin{document}

\title[Instability of nonlinear scalar field on strongly charged AdS BH]{Instability of nonlinear scalar field on strongly charged asymptotically AdS black hole background}

\author{Filip Ficek$^{1,2}$}
\author{Maciej Maliborski$^{1,2,\dagger}$}
\address{$^1$University of Vienna, Faculty of Mathematics, Oskar-Morgenstern-Platz 1, 1090 Vienna, Austria}
\address{$^2$University of Vienna, Gravitational Physics, Boltzmanngasse 5, 1090 Vienna, Austria}
\email[]{filip.ficek@univie.ac.at}
\email[]{maciej.maliborski@univie.ac.at}
\address{$^{\dagger}$Corresponding author}

\begin{abstract}

The conformally invariant scalar equation permits the Robin boundary condition at infinity of asymptotically-AdS spacetimes.
We show how the dynamics of conformal cubic scalar field on the Reissner-Nordstr\"{o}m-anti-de Sitter background depend on the black hole size, charge, and choice of the boundary condition. We study the whole range of admissible charges, including the extremal case. In particular, we observe the transition in stability of the field for large black holes at the specific critical value of the charge. Similarities between Reissner-Nordstr\"{o}m and Kerr black hole let us suspect that a similar effect may also occur in rotating black holes.
	
\end{abstract}

\date{\today}

\maketitle

\tableofcontents

\section{Introduction}
\label{sec:Introduction}

The goal of this work is to investigate nonlinear dynamics of a cubic conformal scalar field propagating on the Reissner-Nordstr\"{o}m-anti-de Sitter (RNAdS) background with Robin boundary condition.
Therefore, it aims to generalise the results of the preceding paper \cite{FFMM24} to charged black holes.
Similarly to \cite{FFMM24} we restrict the analysis to spherically symmetric solutions, and investigate how the dynamics depends on the properties of the black hole (radius and charge) and the Robin boundary parameter.
We find that if the charge is not too big then we observe qualitatively the same behaviour as in \cite{FFMM24}.
Sufficiently large black holes dissipate the scalar field strongly enough
and lead to a nonlinearly stable trivial solution for defocusing nonlinearity and prevent blowup for small data for focusing nonlinearity.

However, for black holes with charge above a certain critical value, we observe a transition to more unstable behaviour.
Even in the regime of large black holes the trivial solution becomes nonlinearly unstable: for the defocusing case we observe convergence to a nontrivial static solution, for the focusing case any perturbation blows up in finite time.
Notably, this transition occurs independently of the boundary condition. 

The described change in stability follows from the emergence of the growing mode in the perturbative analysis around zero solution.
This effect has been recently observed for scalar fields on the near-extremal black hole background with Dirichlet and Neumann boundary conditions \cite{Zheng.2024}.
There, in the presence of the black hole with a sufficiently large charge, the energy functional of the field becomes unbounded from below.
On the other hand, an analogous phenomenon has been known to occur for the fields with a specific Robin boundary condition on the uncharged black hole backgrounds \cite{Holzegel.2014pzh}. 
In this case the instability is driven by the inflow of the energy through the conformal boundary. Our results show that these two seemingly distinct mechanisms coexist in a large part of the parameter space. They also form two regimes between which one can move smoothly.

In addition, we also study the extremal RNAdS black hole. We observe there qualitatively similar behaviour as in the subextremal case. However, depending on the parameters in the model there are noticeable differences. 
In the defocusing case, the global attractor becomes singular at the horizon, thus any initial data leads to singularity formation at infinite time.
For the focusing nonlinearity we also observe  changes with respect to the subextremal black holes.

This paper is structured as follows.
Section \ref{sec:Model} introduces the model.
In Section \ref{sec:LinearEquation}, as an important step before tackling the nonlinear problem, we investigate static solutions of the linearised equation. Such solutions were found in \cite{FFMM24} to divide the parameter phase space into stable and unstable regions.
Subsequently, in Sections \ref{sec:DefocusingNonlinearity} and \ref{sec:FocusingNonlinearity} we discuss nonlinear equation with defocusing and focusing nonlinearities, respectively. 
Both sections are similarly structured, beginning with the discussion of the static solutions and then covering the dynamical behaviour. We end up with conclusions in Section~\ref{sec:Conclusions}.

\section{Model}
\label{sec:Model}

We consider massive cubic wave equation 
\begin{align}\label{eqn:scalar_eqn}
    \Box \phi+2\phi-\lambda \phi^3=0\,,
\end{align}
with either $\lambda=-1$ (focusing case) or $\lambda=1$ (defocusing case) on a fixed background of RNAdS black hole.
The specific choice of the conformal mass (\mbox{$m^{2}=-2$}) in \eqref{eqn:scalar_eqn}
gives us the freedom to prescribe the boundary condition at the conformal boundary \cite{FFMM24}, see also \cite{Bizoń.2014}.
Here we are interested in the class of Robin boundary conditions which are parameterized by $\beta\in(-\pi/2,\pi/2]$ in the following way
\begin{align}
    \label{eqn:RobinBC_0}
    \lim_{r\to\infty}\left(r^2 \partial_r (r\phi) \cos\beta + (r\phi)\sin\beta\right)=0\,.
\end{align}
Note the relation $b=\tan\beta$ with  the parameter $b$ used in \cite{FFMM24}, and that both endpoints $\beta=\pm\pi/2$ can be identified with each other.

Recall that in the Schwarzschild spherical coordinates $(t,r,\theta,\phi)$, the line element of the RNAdS spacetime is given by
\begin{equation*}
    \diff{s}^2=-V(r)\diff{t}^2+V(r)^{-1} \diff{r}^2+r^2 \diff{\Omega}^2\,,
\end{equation*}
where $\diff{\Omega}^2$ is a round metric on a unit two-dimensional sphere and
\begin{align*}
    V(r)=1-\frac{2M}{r}+\frac{Q^2}{r^2}+\frac{r^2}{\ell^2}\, ,
\end{align*}
with $M$ and $Q$ standing for the black hole mass and charge, respectively. The length scale parameter $\ell$ is related to the cosmological constant $\Lambda=-3/\ell^{2}$. In this work we set the units such that $\ell=1$.
In the generic case, this spacetime has two spherical horizons, located at the coordinates $r_{C}$ and $r_{H}$, which are the roots of $V(r)$. These are referred to as the Cauchy horizon (of a radius $r_C$) and the event horizon (of a radius $r_H$). We introduce the parameter $\sigma=r_C/r_H$ as a measure of charge of the black hole: when $\sigma=1$ the horizons coincide and the black hole is extremally charged, while $\sigma=0$ refers to the uncharged case (Schwarzschild-anti-de Sitter spacetime, or SAdS for short), reducing the problem to the one investigated in \cite{FFMM24}. 
Using this notation the metric function $V(r)$ can be written as
\begin{align*}
    V(r)=1-\frac{(1+\sigma)r_H}{r} \left(1+\left(1+\sigma ^2\right)r_H^2 \right)+\frac{\sigma r_H^2}{r^2}\left(1+\left(1+\sigma+\sigma^2\right) r_H^2 \right)+r^2\,.
\end{align*}

Now we introduce the null coordinate $v$ defined by (an alternative way of desingularising the metric would be to use an appropriate hyperboloidal foliation \cite{FFCW24})
\begin{align*}
    \diff{v}=\diff{t}+\frac{\diff{r}}{V(r)}
\end{align*}
and compactify the space by $y=1/r$. It is also convenient to define $y_H=1/r_H$. These coordinate changes reduce Eq.~\eqref{eqn:scalar_eqn} to
\begin{align}\label{eqn:scalar_eqn_2}
    2\partial_y \partial_v \Phi-\partial_y\left(y^2 U(y) \partial_y\Phi\right)-\left(y\, U'(y)+\frac{2}{y^2}\right)\Phi+\lambda \Phi^3=0\,,
\end{align}
where we have introduced
\begin{equation}
    \label{eq:24.09.27_01}
    \Phi(v,y)=r\, \phi(t,r)\,,
\end{equation}
and
\begin{equation}
    \label{eq:24.09.27_02}
    U(y)\equiv V\left(r(y)\right)=\frac{1}{y_H^4 y^2 }(y_H-y) (y_H-\sigma y ) \left(\left(1+y_H^2+\sigma+\sigma ^2\right)y^2+(1+\sigma)y_H \, y+y_H^2\right)\,.
\end{equation}
Moreover, the Robin boundary condition \eqref{eqn:RobinBC_0} transforms into
\begin{align}\label{eqn:RobinBC_1}
    \left.\left[\left(-\partial_v \Phi +\partial_y \Phi\right)\cos \beta -\Phi \sin \beta\right]\right|_{y=0}=0\,.
\end{align}
Note that Eq.\ \eqref{eqn:scalar_eqn_2} has no singularity at $y=0$, since $y^2U(y)=1+\mathcal{O}(y^2)$, but a singularity occurs at the other end $y=y_H$:
\begin{multline}\label{eqn:singularity}
    y^2U(y)=\frac{(1-\sigma)\left(3+y_H^2+2\sigma+\sigma ^2\right)}{y_H}(y_H-y)
    \\
    -\frac{(2-3\sigma)y_H^2+3\left(1-\sigma-\sigma^2-\sigma^3\right)}{y_H^2}(y_H-y)^2+\mathcal{O}\left((y_H-y)^3\right)\,.
\end{multline}

We follow the line of work introduced in \cite{FFMM24} where the key role is played by the static solutions. They can be obtained by putting $\Phi(t,y)=u(y)$ in Eq.~\eqref{eqn:scalar_eqn_2} so one gets
\begin{align}\label{eqn:static}
    -\left(y^2 U(y)\, u'(y) \right)'-\left(y\, U'(y)+\frac{2}{y^2}\right) u(y)+\lambda\, u(y)^3=0\,,
\end{align}
where the prime denotes the $y$ derivative.
The aforementioned singularity at the horizon requires that a static solution with some $u(y_H)=c$ to be regular must satisfy (for $\sigma<1$)
\begin{align}\label{eqn:initial}
    u'(y_H)=-\frac{1-\sigma-\sigma^2-\sigma^3+y_H^2(1-\sigma)+\lambda y_H^2 c^2}{y_H (1-\sigma)(3+y_H^2+2\sigma+\sigma^2)}\,c\, .
\end{align}
This condition together with Eq.~\eqref{eqn:static} constitutes an ODE initial value problem that can be solved numerically for every finite value of $c$. If the solution extends to $y=0$, one can read the value of the Robin parameter from Eq.~\eqref{eqn:RobinBC_1}. Hence, we get the correspondence between the values of $c$ and $\beta$ that will be explored in the following sections.

In the extremal case ($\sigma=1$) the first term in \eqref{eqn:singularity} vanishes, so the qualitative behaviour changes. As a result, it is no longer possible to find non-trivial regular solutions to Eq.~\eqref{eqn:static}, regardless of the value of $\lambda$.
To see this, note that in Eq.~\eqref{eqn:initial} one needs $c^2=2/\lambda y_H^2$ in order to compensate for the zero in the denominator.
For linear ($\lambda=0$) and focusing ($\lambda=-1$) equations this condition cannot hold, while in defocusing case ($\lambda=1$) it leads to the explicit solution $u(y)=\sqrt{2}/y$ that is singular at $y=0$.
In spite of this, static solutions with singularity at the horizon may still exist and play a role in the dynamics of the system.
This issue will be discussed in a greater detail below.

\section{Linear equation}
\label{sec:LinearEquation}

We linearise \eqref{eqn:scalar_eqn_2} around $\Phi=0$ and introduce separation of variables
\begin{equation}
    \label{eq:24.09.26_01}
    \Phi(v,y) = \varepsilon e^{i\omega v}u(y)
    \,,
    \quad
    |\varepsilon|\ll 1\,.
\end{equation}
Then we obtain
\begin{equation}
    \label{eq:24.09.26_02}
    2i\omega u'(y) = \left(y^2 U(y)\, u'(y) \right)'+\left(y\, U'(y)+\frac{2}{y^2}\right) u(y)\,.
\end{equation}
Solutions to this equation satisfying regularity condition at the horizon and the Robin boundary condition can be understood as quasinormal modes (QNM) with frequency $\omega$.
Depending on the sign of the imaginary part of $\omega$ we can infer about linear stability of the trivial solution.
QNMs with Dirichlet boundary condition were already computed in \cite{FFCW24} using the Leaver approach. The method described there could be extended to also find QNM with Robin boundary conditions. Alternatively they can be obtained either with shooting method or the algebraic approach, as done in \cite{FFMM24} for the uncharged case. Yet another option would be to use the Horowitz–Hubeny method, see \cite{Berti.2003bw9} and references therein.

\begin{figure}[!t]
    \centering
    \includegraphics[width=0.32\linewidth]{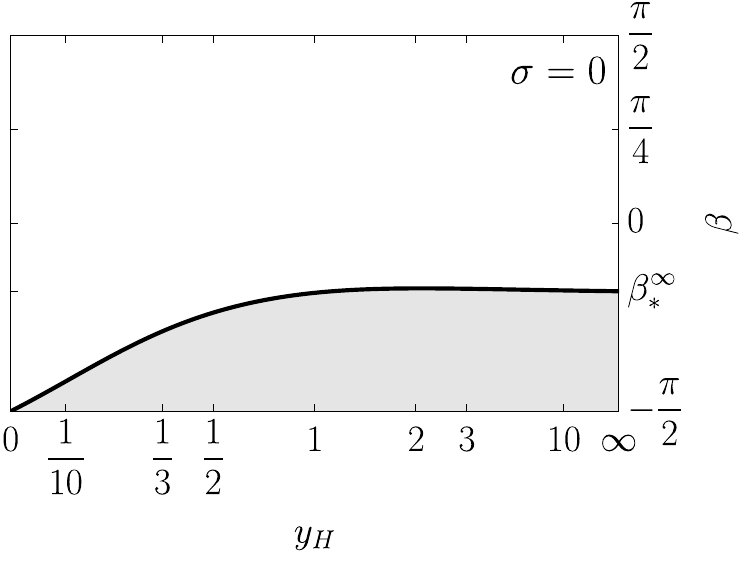}
    \includegraphics[width=0.32\linewidth]{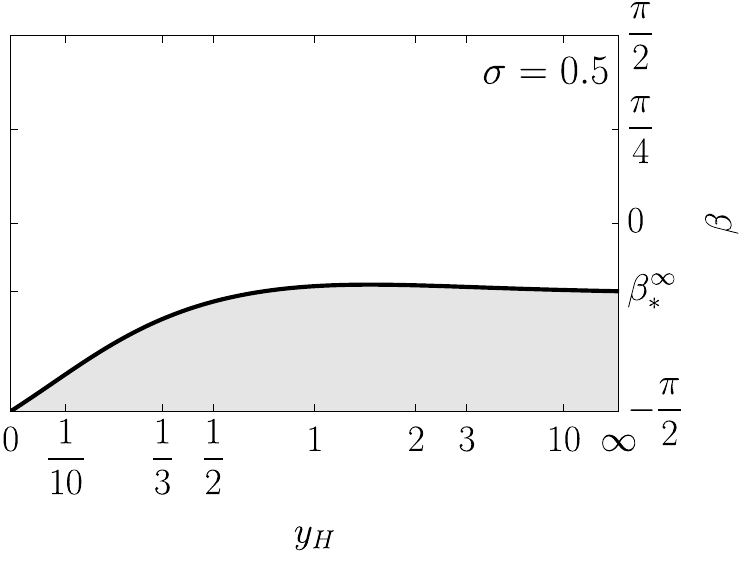}
    \includegraphics[width=0.32\linewidth]{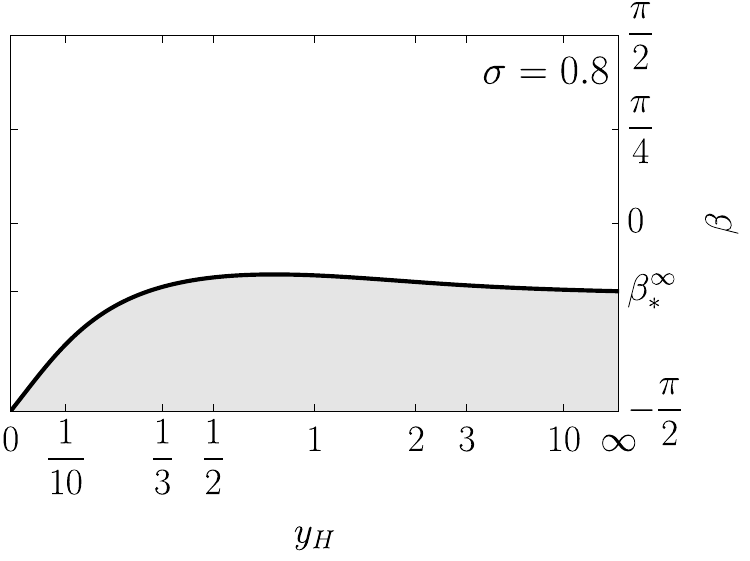}\\
    \includegraphics[width=0.32\linewidth]{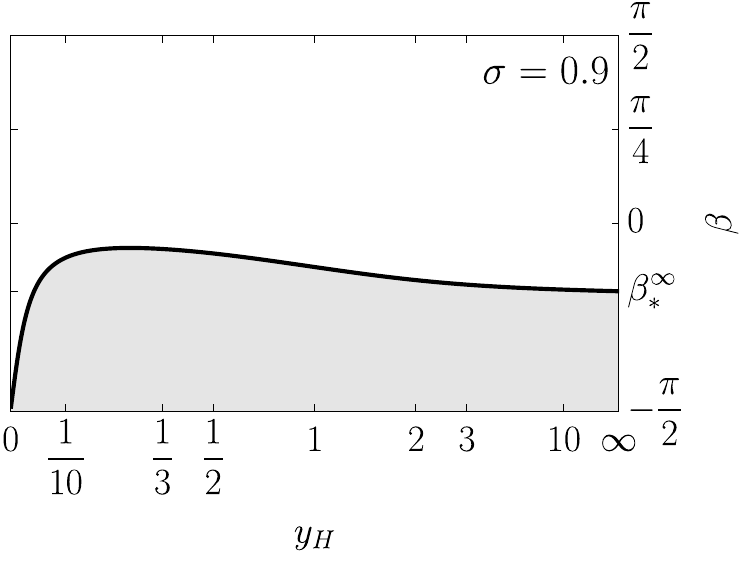}
    \includegraphics[width=0.32\linewidth]{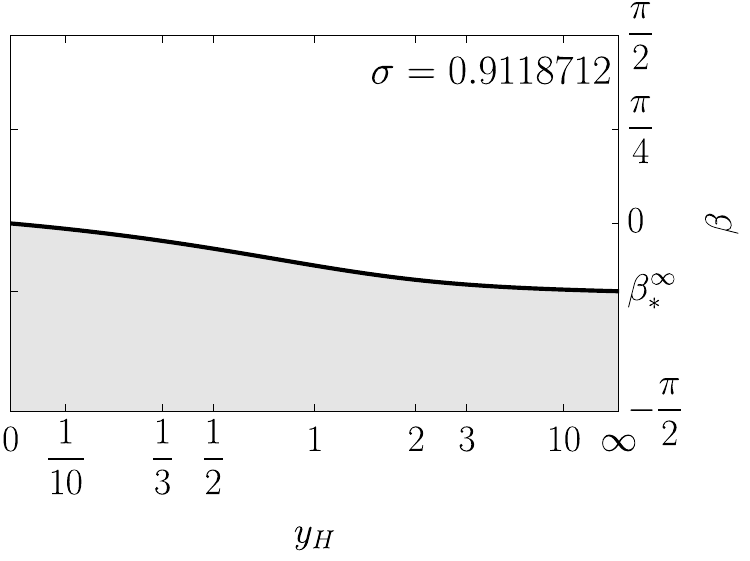}
    \includegraphics[width=0.32\linewidth]{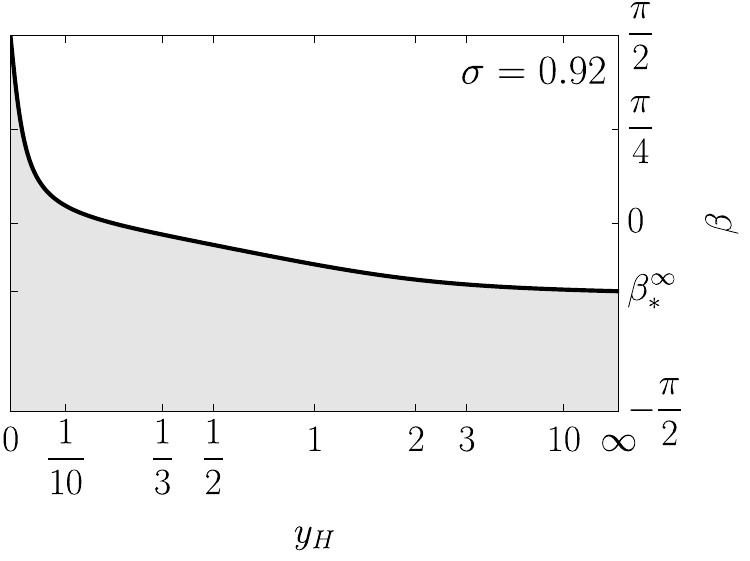}
    \caption{Phase plots $(y_{H},\beta)$ for increasing charge $\sigma$. For $\beta$ below the critical curve $\beta_*(y_H,\sigma)$ (black line) the trivial solution $\phi=0$ is linearly unstable. This line also separates regions with distinct behaviour for nonlinear problem.}
    \label{fig:phase_plots_small}
\end{figure}

\begin{figure}
    \centering
    \includegraphics[width=0.32\linewidth]{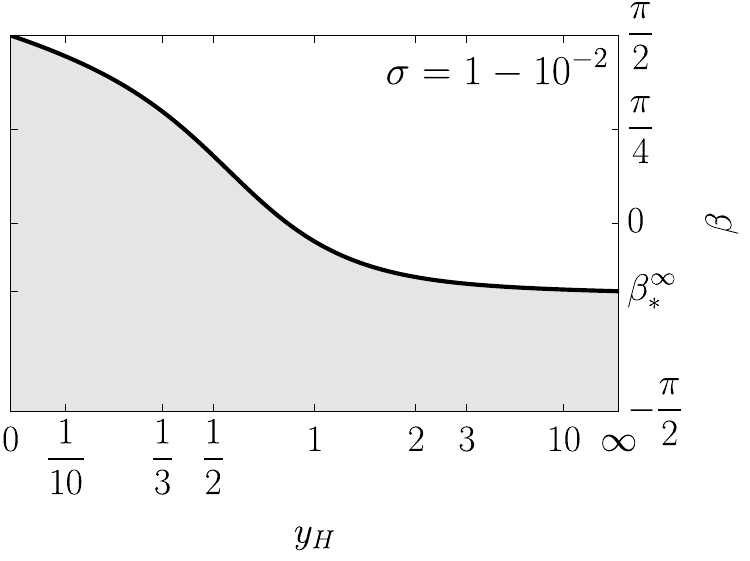}
    \includegraphics[width=0.32\linewidth]{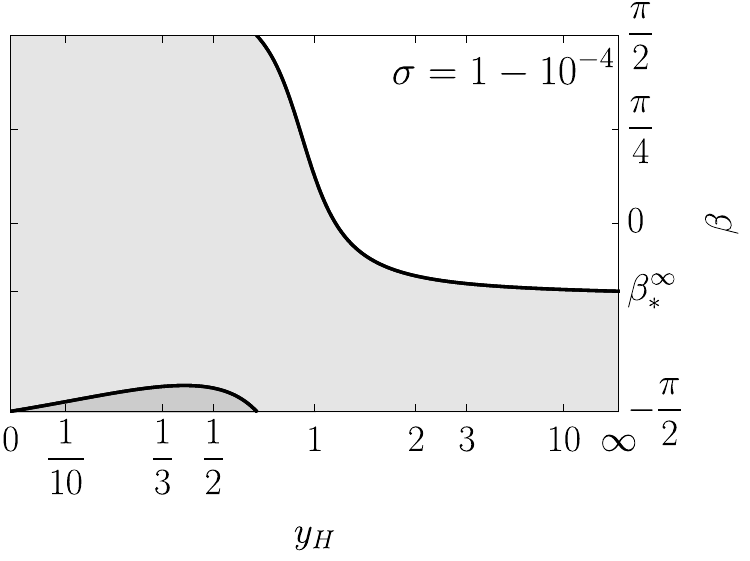}
    \includegraphics[width=0.32\linewidth]{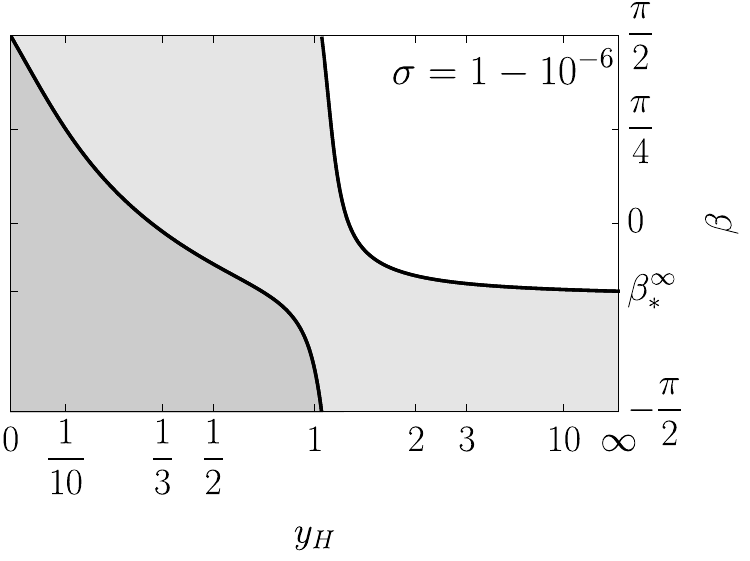}\\
    \includegraphics[width=0.32\linewidth]{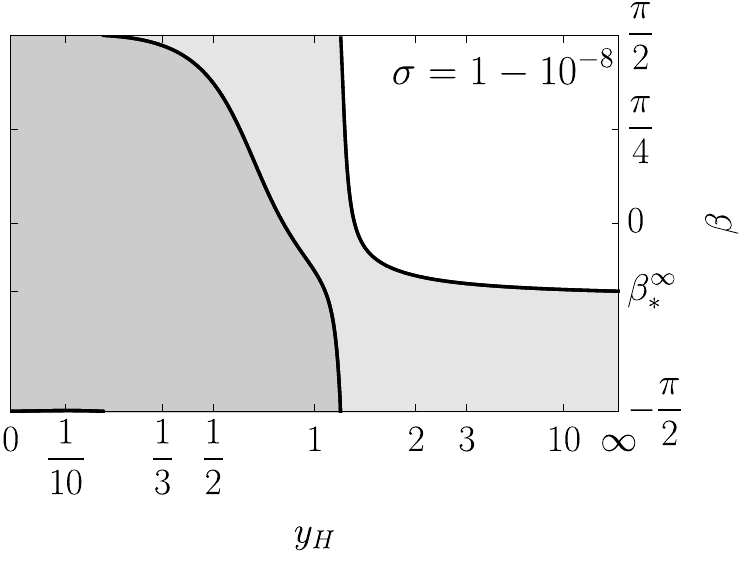}
    \includegraphics[width=0.32\linewidth]{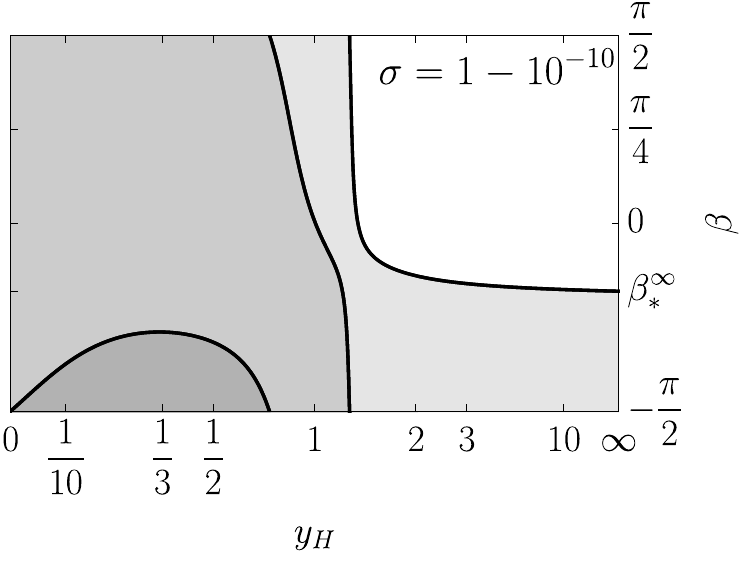}
    \includegraphics[width=0.32\linewidth]{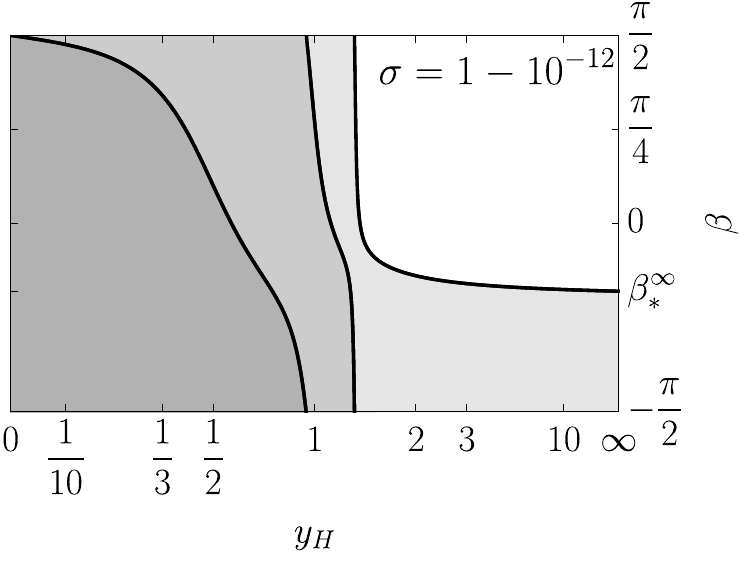}\\
    \includegraphics[width=0.32\linewidth]{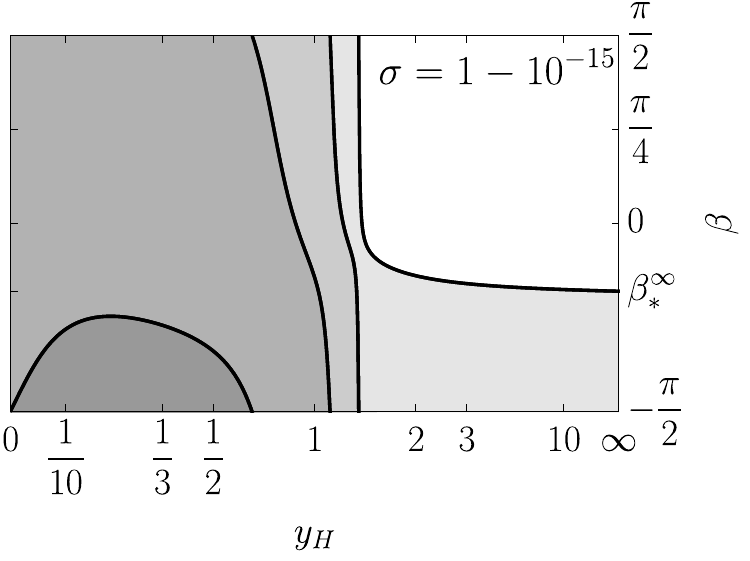}
    \includegraphics[width=0.32\linewidth]{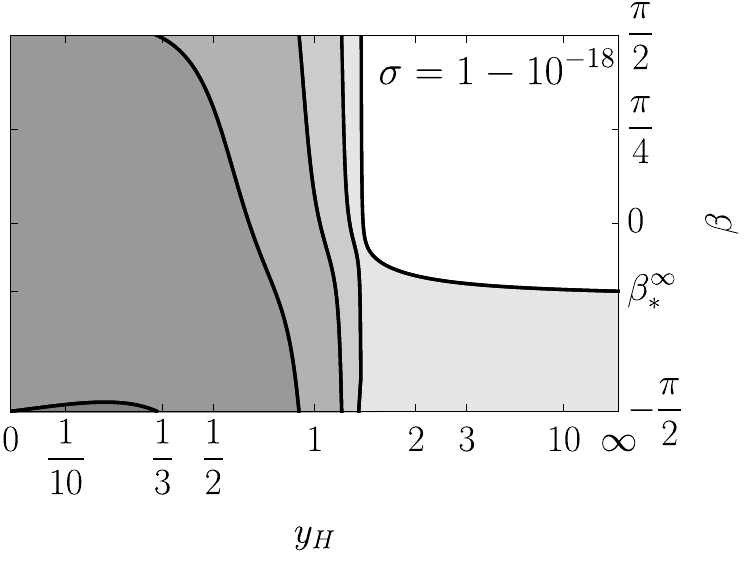}
    \includegraphics[width=0.32\linewidth]{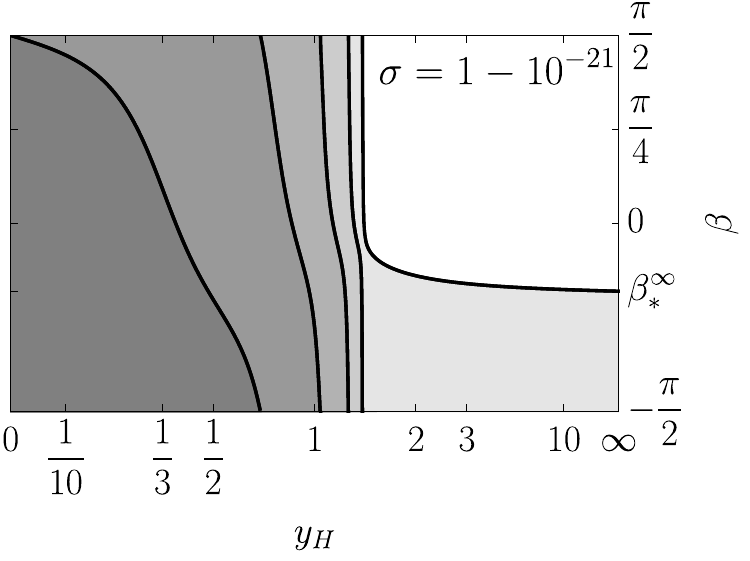}\\
    \caption{
    Phase plots for $\sigma$ approaching the extremality, cf. Fig.~\ref{fig:phase_plots_small}. The critical curve separates the phase space into regions (denoted by various shades of gray) characterised by different sets of static solutions of the nonlinear equation, see Sec.~\ref{sec:DefocusingStaticSolutions} and \ref{sec:FocusingStaticSolutions}. New regions emerge at charges $\sigma_{w}^{(n)}$, see Tab.~\ref{tab:critical_sigmas}. For $y_{H}<\sqrt{2}$ the winding number goes to infinity as $\sigma\rightarrow 1$.}
    \label{fig:phase_plots_large}
\end{figure}

In this section we focus on static solutions to the linearised equation, hence, we put $\omega=0$ in Eq.~\eqref{eq:24.09.26_02}. As a result we get \eqref{eqn:static} with $\lambda=0$. For $\sigma<1$ this equation has two families of independent solutions, one regular at the horizon $y=y_H$ and one singular, behaving as $\log(y_H-y)$. Getting the former requires Eq.~\eqref{eqn:initial} to hold.
Since the equation is linear, a specific choice of $c$ does not affect the Robin parameter $\beta$.
Hence, to each $y_H>0$ and $0\leq \sigma<1$ one can associate a unique $\beta=\beta_\ast(y_H,\sigma)$. We present these values as black lines, called critical curves, on phase space plots, see Figs.\ \ref{fig:phase_plots_small} and \ref{fig:phase_plots_large}. The same argument as presented in \cite{FFMM24} shows that regardless of the choice of $\sigma$ one has $\lim_{y_H\to\infty} \beta_\ast(y_H,\sigma) = \beta_\ast^\infty:=-\arctan{(2/\pi)}$. In the gray regions below the critical curve the trivial solution is linearly unstable, see also \cite{Zheng.2024} for similar result for charged scalar field.

Plots in Fig.~\ref{fig:phase_plots_small} show the phase planes for charges $\sigma\leq 0.92$. For $\sigma=0$ we get the same plot as presented in \cite{FFMM24,Holzegel.2014pzh} (the differences come from the compactification). As the charge $\sigma$ increases, the plots initially look qualitatively the same. However, for $\sigma^{(1)}_{c}=0.9118712...$ one observes that in the limit of $y_H\to 0$ the linear solution satisfies the Neumann boundary condition ($\beta=0$) rather than Dirichlet boundary condition ($\beta=\pm\pi/2$).\footnote{As discussed in \cite{FFMM24}, in the limit $y_H\to 0$ (large black holes limit) the solution must satisfy either the Neumann $(\beta=0)$ or Dirichlet $(\beta=\pm\pi/2)$ boundary condition. This result also holds for RNAdS background with $\sigma<1$.} For slightly larger $\sigma$, the solution in the $y_H\to 0$ limit again satisfies the Dirichlet boundary condition, however, the region with small values of $y_H$ is now "below" the critical curve (compare plots for $\sigma=0.9$ and $\sigma=0.92$ in Fig.~\ref{fig:phase_plots_small}). 
This result holds for all $\sigma>\sigma^{(1)}_{c}$ and, as we will see in the following sections, has major implications for the dynamics of the scalar field with non-Dirichlet boundary condition in the presence of large black holes. (For Dirichlet boundary condition analogous transition takes place at $\sigma=\sigma^{(1)}_{w}>\sigma^{(1)}_{c}$ defined below.)
As $\sigma$ increases further, one encounters successive values of charge $\sigma^{(n)}_{c}$ for which again Neumann BC is satisfied in the limit $y_H\to 0$, see Tab.~\ref{tab:critical_sigmas}. However, at those values the generic dynamics of the system does not experience any significant changes.

Discussion of the systems with larger $\sigma$ can be simplified by introducing the notion of the winding number. For any fixed $\sigma$ and $y_H$ we define the winding number as the number of zeroes of the regular static solution to the linearised equation \eqref{eq:24.09.26_02}. In particular, for all parameters covered by Fig.~\ref{fig:phase_plots_small} the winding number is zero. However, as $\sigma$ increases (more precisely, at $\sigma^{(1)}_{w}=0.999329...$) the winding number for small values of $y_H$ changes to one. It can be observed on the phase plots as a passage of the critical curve through the upper boundary of the plot, i.e., through the line $\beta=\pi/2$ corresponding to Dirichlet BC, see Fig.~\ref{fig:phase_plots_large}. As $\sigma$ increases further, higher winding numbers appear at values $\sigma^{(n)}_{w}$ given in Tab.~\ref{tab:critical_sigmas}. 
In general, for any fixed $\sigma$ and $y_H$, the winding number can be determined from the phase plot by counting the number of times the critical curve crosses the upper boundary as one moves from the infinite $y_H$ to the preselected $y_H$.

\begin{table}[]
    \centering
    \begin{tabular}{|c|c|c|}\hline
      \toprule
      $n$ & $1-\sigma^{(n)}_{c}$ & $1-\sigma^{(n)}_{w}$ \\
      \midrule
      1 & $8.813\times 10^{-2}$ & $6.713\times 10^{-4}$ \\
      2 & $1.731\times 10^{-6}$ & $1.261\times 10^{-8}$ \\
      3 & $3.251 \times 10^{-11}$ & $2.368\times 10^{-13}$ \\
      4 & $6.105\times 10^{-16}$ & $4.447\times 10^{-18}$ \\
      5 & $1.147\times 10^{-20}$ & $8.350\times 10^{-23}$ \\
      $\vdots$ & $\vdots$ & $\vdots$ \\
      $n\rightarrow \infty$ & $\exp\left(-10.88 n + 8.473\right)$ & $\exp\left(-10.88 n + 3.577\right)$\\
      \bottomrule
    \end{tabular}
    \vskip 2ex
    \caption{Values of $\sigma^{(n)}_{c}$ and $\sigma^{(n)}_{w}$ charges (see the text for definitions). The asymptotic formula follows from fitting the listed numbers.}
    \label{tab:critical_sigmas}
\end{table}

\begin{figure}
    \centering
    \includegraphics[width=0.7\linewidth]{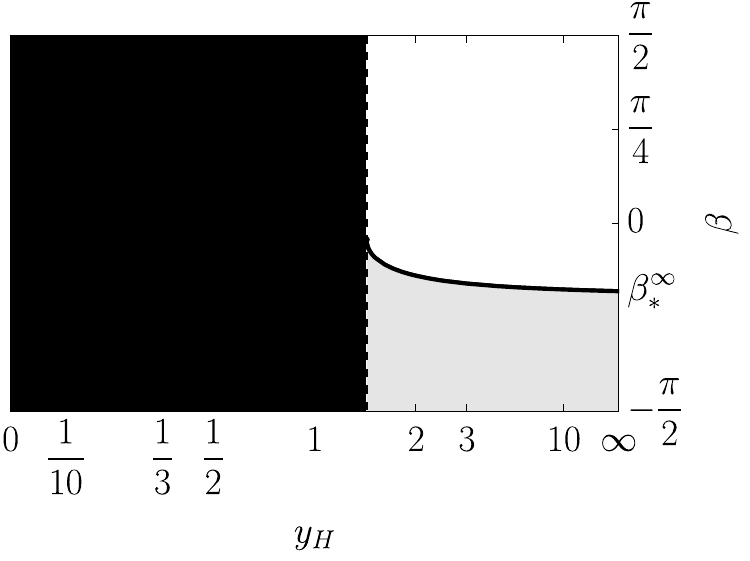}
    \caption{Phase plot for extremal case $\sigma=1$. The critical curve $\beta_{*}(y_{H},1)$ terminates at $y_{H}=\sqrt{2}$. The white and gray regions correspond to the analogous parts of the phase plots shown in Figs.~\ref{fig:phase_plots_small} and \ref{fig:phase_plots_large}, whereas the black part is the unique feature of the extremal case.}
    \label{fig:phase_plots_extremal}
\end{figure}

\begin{figure}
    \centering
    \includegraphics[width=0.75\linewidth]{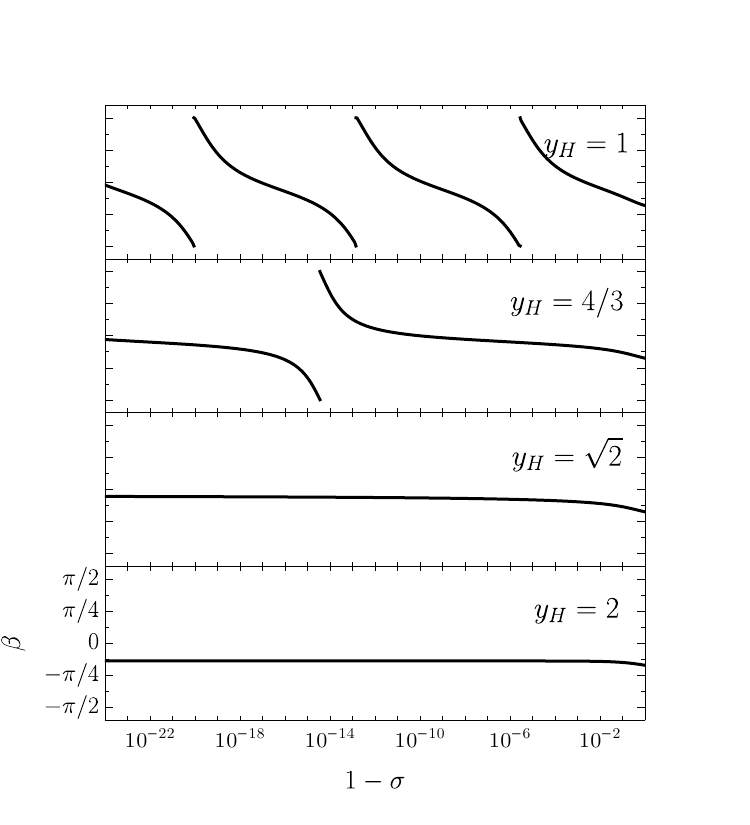}
    \caption{Values of $\beta$ as a function of $1-\sigma$ for various $y_H$. For $y_H\geq\sqrt{2}$ we observe convergence with $(1-\sigma)\rightarrow 0$, whereas for $y_H<\sqrt{2}$ the Robin parameter $\beta$ appears approximately periodic in $\log(1-\sigma)$.}
    \label{fig:beta}
\end{figure}

For $y_H\geq \sqrt{2}$  the critical curve  converges to the shape presented in Fig.~\ref{fig:phase_plots_extremal} in the limit of extremality ($\sigma\to 1$), while for any $y_H<\sqrt{2}$ there is no convergence, see Fig.~\ref{fig:beta}.
In particular, for any fixed $y_H<\sqrt{2}$ as $\sigma\to 1$ the winding number increases steadily with $\log (1-\sigma)$, as can be seen in two first plots of Fig.~\ref{fig:beta}.
This difference in the behaviours can be explained by investigation of the extremal case.
For $\sigma=1$ Eq.~\eqref{eqn:static} with $\lambda=0$ has no regular solutions, as discussed above.
To deal with this problem we can factorise the singular behaviour at $y=y_H$ using a Frobenius-like approach. By fixing $u(y)=(y_H-y)^a\, w(y)$ we get the indicial polynomial
\begin{align*}
    a^2+a+\frac{2}{6+y_H^2}
\end{align*}
with the roots given by
\begin{align*}
    a_\pm =\frac{1}{2}\left(-1\pm\sqrt{\frac{y_H^2-2}{y_H^2+6}}\right)\,.
\end{align*}
For $y_H\geq \sqrt{2}$ both of the roots are real and we can focus on the less singular branch by picking $a_+$. It leads to a well-posed initial value problem for $w(y)$, 
that can be solved analogously to the subextremal case ($\sigma<1$). Its solution is a limit for subextremal cases with fixed $y_H\geq \sqrt{2}$ as $\sigma\to 1$. It gives us the right part of the phase plot shown in Fig.~\ref{fig:phase_plots_extremal} (the boundary between white and gray regions). 
On the other hand, there is no such convergence for $y_H<\sqrt{2}$, as the roots of the indicial polynomial are complex. Thus, for $\sigma=1$ any solution would consist of infinitely many oscillations with unbounded amplitude in the vicinity of the black hole horizon.

\section{Defocusing nonlinearity}
\label{sec:DefocusingNonlinearity}

\subsection{Static solutions}
\label{sec:DefocusingStaticSolutions}

\begin{figure}
    \centering
    \includegraphics[width=0.45\linewidth]{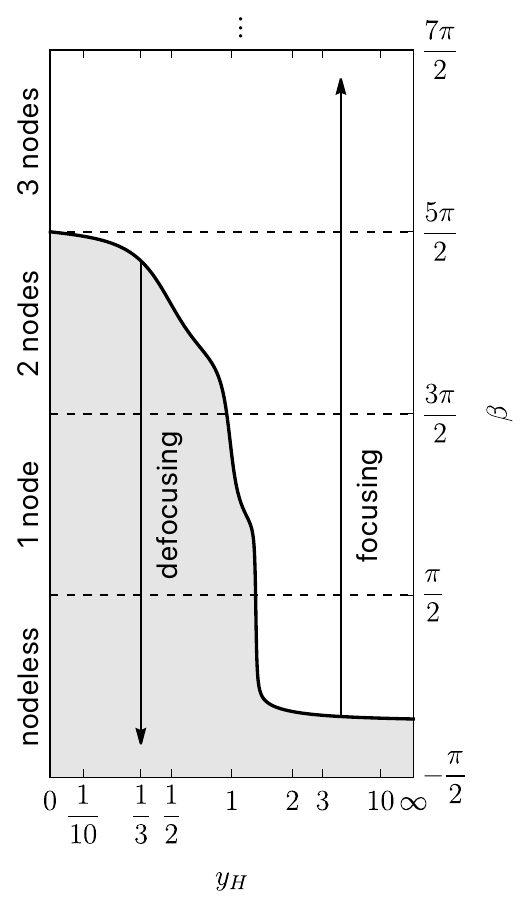}
    \caption{Phase space for $\sigma=1-10^{-12}$ without the periodic identification of the Robin parameter $\beta$. Each point of this diagram represents a static solution. Static solutions for parameters $(y_{H},\beta)$ on the dashed lines satisfy Dirichlet boundary condition. Those lines separate regions populated by static solutions with different numbers of zeroes.
    Solutions with small amplitude $c$, see \eqref{eqn:static} and \eqref{eqn:initial}, bifurcate from $u=0$ at the critical curve $\beta_{*}$ in directions presented by the arrows, depending on the type of the nonlinearity. Hence, the critical curve $\beta_{*}$ separates the diagram into two regions, gray and white, in which there exist static solutions to the defocusing and focusing cases respectively. As $c$ increases static solutions in the defocusing case eventually become singular at $\beta=-\pi/2$, while for the focusing nonlinearity the Robin parameter $\beta$ increases indefinitely and static solutions with more and more zeroes appear. The original phase plot, cf.~Fig.~\ref{fig:phase_plots_extremal}, can be recovered by cutting this plot along the dashed lines and stacking the obtained fragments one on another.}
    \label{fig:unwinded}
\end{figure}

Structure of the static solutions to the defocusing equation can be easily understood by setting $\sigma$ and $y_H$ and solving ODE \eqref{eqn:static}-\eqref{eqn:initial} with $\lambda=1$ for a range of $c$.
At $c=0$ the static solution bifurcates from the linearised solution discussed in Sec.~\ref{sec:LinearEquation} and is characterised by Robin parameter $\beta=\beta_{*}(y_{H},\sigma)$ that can be read out from the phase space plot. As $c$ increases, the angle $\beta$ decreases, see the gray part of Fig.~\ref{fig:unwinded}.
If the winding number for the fixed $\sigma$ and $y_H$ is equal to zero, as $c$ increases the static solution stays nodeless and eventually it blows up at $y=0$ for some finite value of $c$ (in this limit $\beta\to -\pi/2$).
If the winding number at the bifurcation is non-zero, as $c$ increases the static solution unwinds passing the Dirichlet BC lines and it loses zeroes, eventually becoming nodeless.
As we increase $c$ even further, the solution shares the fate of the previous case (starting with zero winding number) and blows up at some finite $c$.
Let us point out that the nodeless static solutions exists only in the gray regions of the phase plots.

\begin{figure}
    \centering
    \includegraphics[width=0.45\linewidth]{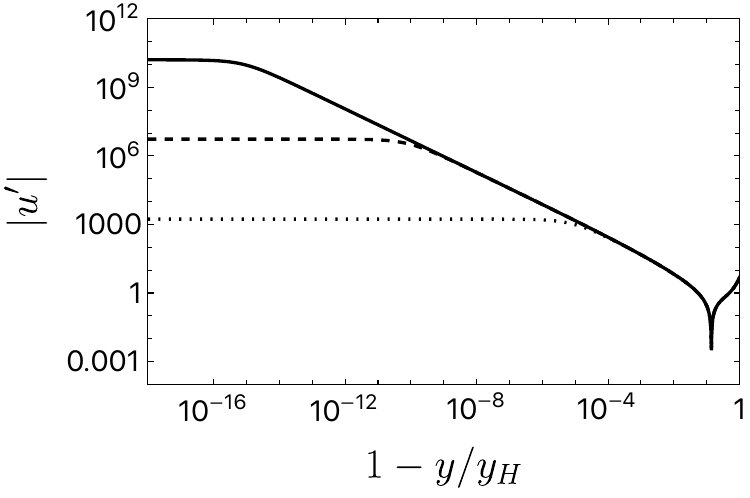}
    \includegraphics[width=0.45\linewidth]{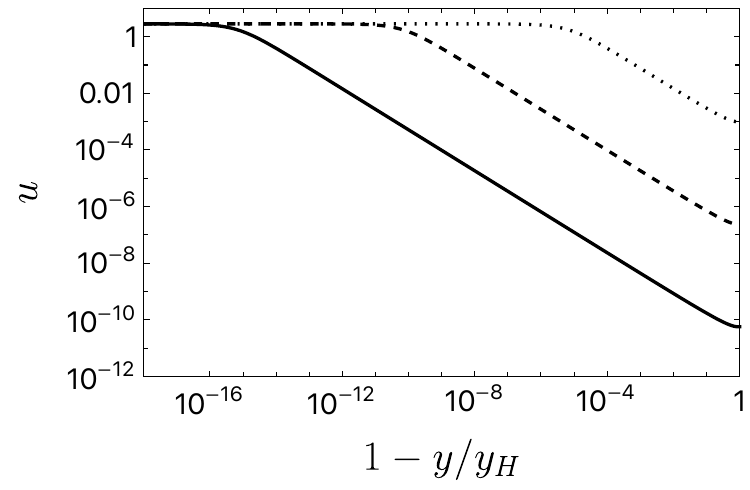}

    \vspace{1ex}

    \includegraphics[width=0.80\linewidth]{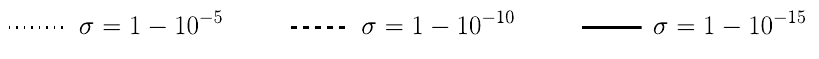}
    
    \caption{Changes in the profiles of nonlinear static solutions as $\sigma\rightarrow 1$. The left plot presents absolute values of derivatives $u'$ for defocusing case with $y_H=2$ and $\beta=-\arctan 1$. The right plot shows solutions $u$ for focusing case with $y_H=2$ and $\beta=0$.}
    \label{fig:sigma_convergence}
\end{figure}

For $\sigma$ approaching one with any fixed $y_H<\sqrt{2}$ and $\beta$, static solutions with more zeroes appear. On the other hand, for $y_H>\sqrt{2}$ the situation is different. If $\beta$ is such that the parameters lie under the critical line in Fig.~\ref{fig:phase_plots_extremal}, there exists a unique static solution and it is nodeless. Above this line there are no static solutions.
Solutions in the gray regions get infinitely steep in the vicinity of the horizon $y=y_H$ as $\sigma\to 1$.
It is in agreement with the fact that in the extremal case ($\sigma=1$) there are no regular static solutions, as follows from the analysis in Sec.~\ref{sec:Introduction}.
The limiting case can be investigated by factoring out the singular behaviour 
\begin{equation}
    u(y)=\sqrt{2}+\sqrt{2}(y_H-y)+(y_H-y)^\gamma v(z),
\end{equation}
where
\begin{equation}
    \gamma=\frac{1}{2}\left(-1+\sqrt{\frac{y_H^2+22}{y_H^2+6}}\right), \qquad z=(y_H-y)^\gamma.
\end{equation}
Using these new variables one gets standard initial value problem at $y=y_H$ that can be solved numerically. As a result, for $y_H<\sqrt{2}$ (black region of Fig.~\ref{fig:phase_plots_extremal}) there exists the whole ladder of the static solutions beginning with the nodeless one, while for $y_H\geq\sqrt{2}$ there are no static solutions for $\beta$ above the critical curve (white region) and there is a single, nodeless static solutions if $\beta<\beta_\ast$ (gray region). 
Solutions $u(y)$ are bounded but their first derivatives blow up to infinity as $y\to y_H$, see Fig.~\ref{fig:sigma_convergence}.

\subsection{Dynamics}
\label{sec:DefocusingDynamics}
For a subextremal RNAdS black hole any regular initial data solutions to the defocusing equation enjoy global-in-time existence.\footnote{To solve the initial-boundary value problem \eqref{eqn:scalar_eqn_2}-\eqref{eqn:RobinBC_1} we essentially follow the approach described in \cite{FFMM24}.
Since the extreme cases, $\sigma$ very close to or equal to one, require high resolution in the vicinity of the black hole,  
we use a new radial coordinate, defined as
\mbox{$z=-1+2\left(y_{M}+\log(e^{-y_{M}}+1-{y}/{y_H})\right)/\left(y_{M}+\log(1+e^{-y_{M}})\right)$}, with a parameter $y_{M}>0$ fixed such that the numerical grid allows us to resolve the fine details in the solution. 
In addition, we use exponential time $v=\exp \nu$.
}
Analogously to the SAdS case \cite{FFMM24}, in the absence of the nodeless static solution (white regions on the phase plots) the trivial solution $\phi=0$ is nonlinearly stable and any initial data configuration eventually converges to it.
However, at a segment of the critical curve $\beta_{*}(\sigma,y_H)$ separating the white and gray regions one observes a pitchfork bifurcation: a pair of stable static solutions\footnote{The linear stability analysis of the (nonlinear) static solutions can be carried out using the methods discussed in \cite{FFMM24}.} (differing only in sign) emerges and $\phi=0$ becomes unstable. Those nodeless static solutions play the role of the global attractors of the evolution in the gray regions of the phase plots.

\begin{figure}
    \centering
    \includegraphics[width=0.496\linewidth]{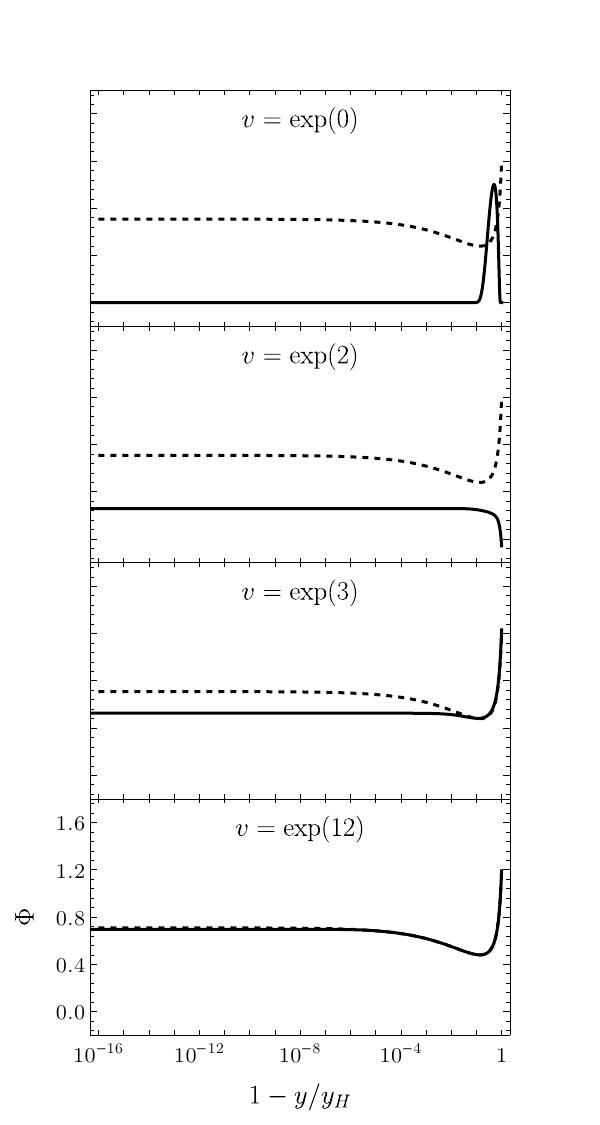}
    \includegraphics[width=0.496\linewidth]{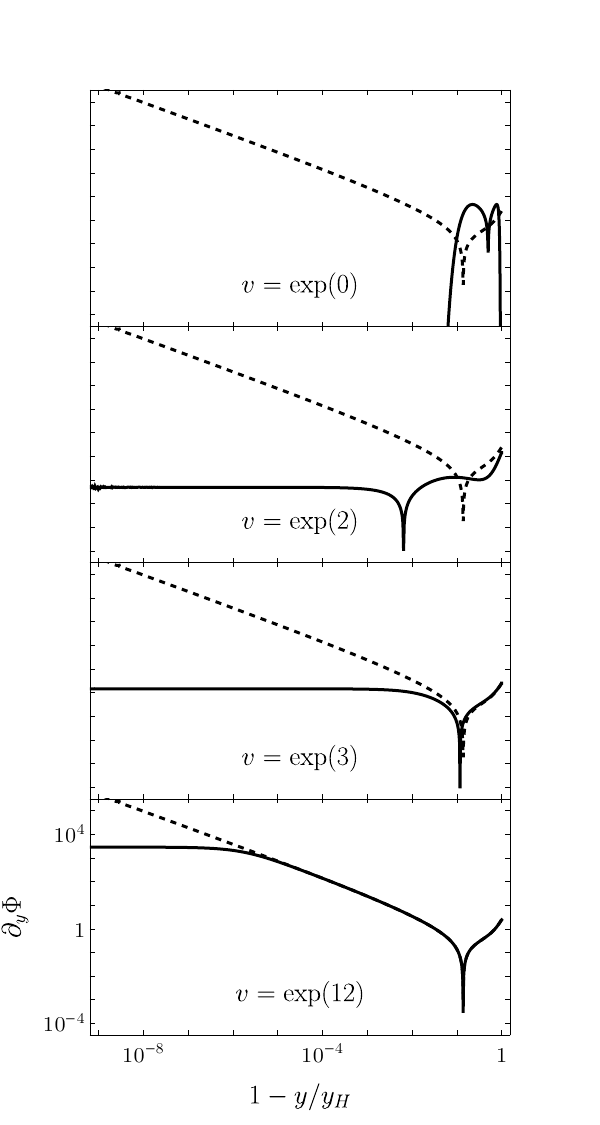}
    \caption{Snapshots from the evolution of a smooth initial data for extremal case and defocusing nonlinearity. At later times we observe convergence to the (singular) static solution. As a result, the gradient of the solution at the horizon $y=y_{H}$ diverges at $v=\infty$.}
    \label{fig:evolution_defocusing_extremal}
\end{figure}

A similar result holds for the extremal case. In the white part of Fig.~\ref{fig:phase_plots_extremal} we observe convergence to the zero solution, and in the gray and black regions nodeless static solutions act as attractors. Since those static solutions are singular at the horizon, as described above, the gradient at the horizon diverges as $v\rightarrow\infty$; see Fig~\ref{fig:evolution_defocusing_extremal}.

\section{Focusing nonlinearity}
\label{sec:FocusingNonlinearity}

\subsection{Static solutions}
\label{sec:FocusingStaticSolutions}
The analysis of static solutions to the focusing ($\lambda=-1$) equation is similar to the other type of nonlinearity.
In this case, we also observe bifurcation from the solutions of the linearised equation, but here, $\beta$ increases with $c$; see Fig.~\ref{fig:unwinded}.
It means that the static solutions wind up, acquiring a new zero with each passage through the Dirichlet boundary condition. 
Consequently, static solutions exist for every value of $c$,
and in the white regions of the parameter plots, there is a whole ladder of static solutions enumerated by the number of nodes, starting from the nodeless solution.
In the gray areas, the nodeless solutions, and potentially some excited solutions, are lost: the darker the area, the more lower-order solutions are absent.

In the extremal case, there are no nontrivial static solutions for the focusing nonlinearity.
When $y_H<\sqrt{2}$ it follows from the fact that for any $(y_H,\beta)$ all static solutions with a fixed number of zeroes eventually cease to exist as $\sigma\to 1$ (because the critical curve ``moves to the right'' of the phase space and accumulates on the vertical line $y_H=\sqrt{2}$).
On the other hand, for $y_H\geq\sqrt{2}$ static solutions for $\sigma<1$  converge pointwise to zero on the interval $y\in[0,y_H)$ as $\sigma\to 1$; see Fig.~\ref{fig:sigma_convergence}.

\subsection{Dynamics}
\label{sec:FocusingDynamics}

\begin{figure}
    \centering
    \includegraphics[width=0.9\linewidth]{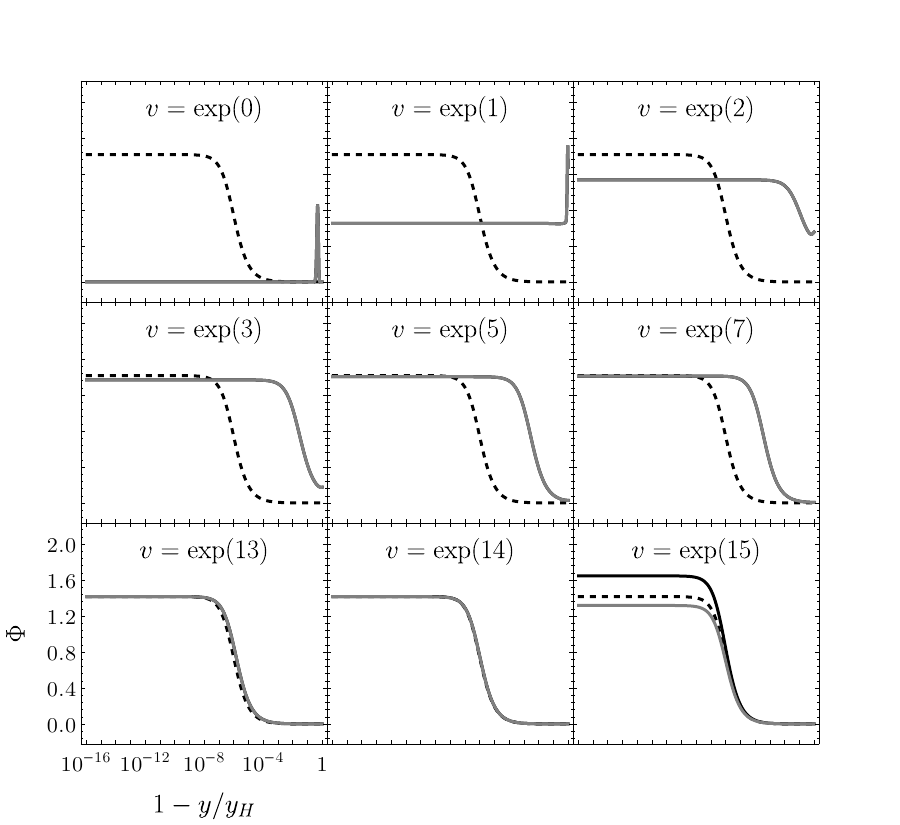}
    \caption{Snapshots of the near-critical evolution for the nearly extremal case $\sigma=1-10^{-6}$ and focusing nonlinearity. Shown are the results for $y_{H}=2$ and $\beta=0$, a representative point in the white region of the phase plane; c.f. Fig.~\ref{fig:phase_plots_large}. The initial data was fine-tuned to the threshold with double machine precision. The supercritical (black) and subcritical (gray) solutions are indistinguishable as they approach the static solution (dashed). However, at later times, the two solutions diverge due to the admixture of an unstable mode, with one solution blowing up in finite time and the other dispersing to zero.}
    \label{fig:FocusingDynamics}
\end{figure}

Similarly to the SAdS case \cite{FFMM24}, for the parameters from the gray regions, where the co-dimension one stable nodeless static solution is absent, we observe a finite-time ODE type blow-up for any nontrivial initial data (the existing static solutions have higher number of unstable modes, equal to the number of nodes, and thus they do not play a role in the evolution starting from generic initial data).
However, in the white regions of the phase plots, where nodeless static solutions exist, they play the role of the critical solution separating two types of behaviour: finite-time blow-up for initial data above the threshold and convergence to zero below it. We illustrate the near critical evolution in Fig.~\ref{fig:FocusingDynamics}, for a particular choice of initial condition
\begin{equation}
    \label{eq:initial_data}
    \Phi_{0}(y) = A \exp\left(16-\frac{y_{H}^4}{y^2 (y-y_{H})^2}\right)
    \,,
\end{equation}
with $A=A_*\pm \Delta A$, $A_{*}\approx 0.864599$, and $\Delta A \sim 10^{-16}$.

This kind of behaviour holds for any subextremal case. However, if we approach the extremality, fine tuning to the blowup threshold becomes more and more difficult. First, the static solution becomes increasingly steep at the black hole horizon, requiring higher and higher resolution there. Secondly, the static solutions possess slowly decaying stable modes and their presence disguises the critical solution for any data tuned with finite precision.
For this reason we did not observe the solution to settle down on the static solution for $\sigma \geq 1-10^{-7}$ when using the double machine precision in the numerical calculations, but only the moving front approaching it and eventually diverging to infinity or dispersing to zero.

Based on the results for $\sigma<1$, we conjecture that, for the extremal case $\sigma=1$, there is still a threshold between dispersion to zero and finite-time ODE-type blowup in the white regions.
However, the critical solution becomes singular as $v\rightarrow\infty$ (the gradient at the horizon becomes unbounded), which is consistent with behaviour of static solutions in the $\sigma\rightarrow 1$ limit (for subextremal case $\sigma<1$ the gradient can become arbitrarily large, but it stays bounded).
This scenario is similar to the behaviour observed in the defocusing case.
In the gray and black regions all solutions blow up in finite time.

\section{Conclusions}
\label{sec:Conclusions}
In the regime of small charges, the dynamics of the conformal field in RNAdS black hole spacetime does not differ significantly from the SAdS case studied in \cite{FFMM24}.
However, as charge gets closer to its extremal value, at $\sigma^{(1)}_{c}=0.9118712...$ (for Dirichlet boundary condition at $\sigma^{(1)}_{w}=0.999329...$), the qualitative behaviour of the field in the presence of the large black hole changes.
Below this critical charge it is more stable:
in the focusing case, there is a threshold separating global existence from finite-time blowup, while in the defocusing case, the zero solution remains nonlinearly stable for arbitrary large data.
Above the critical charge, any non-trivial initial data leads to blow-up in the focusing case, and in the defocusing case, the trivial solution loses stability in favor of a nodeless static solution.

Due to similarities between structures of Reissner-Nordstr\"{o}m and Kerr black holes \cite{PhysRevD.41.1796}, one can suspect that a similar transition occurs also for the system with a rapidly rotating black hole.
However, in such a setup, one cannot expect the scalar field configuration to be spherically symmetric. Lack of this symmetry may lead to much more complicated behaviour of the field \cite{Holzegel.2014}.
Hence, the natural extension of the current work would be the investigation of axially symmetric solutions.
We plan to pursue this goal in the future as a midstep before focusing on the dynamics of the nonlinear field on the Kerr--Anti-de Sitter background. 

Another interesting direction for future research involves the introduction of self-gravitation to our model, thus generalising the previous studies \cite{bizon2020ads, hertog2005designer} to the black hole case.
It has already been proposed in \cite{FFMM24}; however, the current work gives us additional motivation to also consider charged matter, since the presence of the charge may significantly influence the dynamics of the system.

\vspace{4ex}
\noindent\textit{Acknowledgement.} We are grateful to Claude Warnick and Dejan Gajic for helpful remarks. We acknowledge the support of the Austrian Science Fund (FWF), Project \href{http://doi.org/10.55776/P36455}{P 36455} and the START-Project \href{http://doi.org/10.55776/Y963}{Y963}.

\printbibliography

\end{document}